\documentclass[aps,prl,reprint,groupedaddress]{revtex4-1}
\usepackage{graphicx} % Include figure files
\usepackage{dcolumn} % Align table columns on decimal point
\usepackage{bm}% bold math
\usepackage{hyperref}% add hypertext capabilities

\begin{document}

\title{Particle production in antiproton induced nuclear reactions}

\author{Zhao-Qing Feng$^{1,2}$}
\email{fengzhq@impcas.ac.cn}
\author{Horst Lenske$^{3}$}
\affiliation{$^{1}$Institute of Modern Physics, Chinese Academy of Sciences, Lanzhou 730000, People's Republic of China        \\
$^{2}$State Key Laboratory of Theoretical Physics and Kavli Institute for Theoretical Physics China, Chinese Academy of Sciences, Beijing 100190, People's Republic of China        \\
$^{3}$Institut f\"{u}r Theoretische Physik der Universit\"{a}t, Giessen 35392, Germany}

\date{\today}

\begin{abstract}
The quantum molecular dynamics model has been improved to investigate the reaction dynamics induced by antiprotons. The reaction channels of elastic scattering, annihilation, charge exchange and inelastic collisions have been included in the model. Dynamics on particle production, in particular pions, kaons, antikaons and hyperons, is investigated in collisions of $\overline{p}$ on $^{12}$C, $^{20}$Ne, $^{40}$Ca, $^{112}$Sn, $^{181}$Ta, $^{197}$Au and $^{238}$U from a low to high incident momentum. The rapidity and momentum distributions of $\pi^{+}$ and protons from the LEAR measurements can be well reproduced. The impacts of system size and incident momentum on particle emissions are investigated from the inclusive spectra, transverse momentum and rapidity distributions. It is found that the annihilations of $\overline{p}$ on nucleons are of importance on the particle production. Hyperons are mainly produced via meson induced reactions on nucleons and strangeness exchange collisions when the incident momentum is below the threshold value of annihilation reaction. A higher nuclear temperature is obtained from the kaon emission, but it has a lower value for hyperon production.

\begin{description}
\item[PACS number(s)]
25.43.+t, 24.10.-i, 24.10.Lx
\end{description}
\end{abstract}

\maketitle

\section{I. Introduction}

The dynamics of antiproton-nucleus collisions is a complex process, which is associated with the mean-field potentials of antinucleons and produced particles in nuclear medium, and also with a number of reaction channels, i.e., the annihilation channels, charge-exchange reaction, elastic and inelastic collisions. A more localized energy deposition is able to be produced in antiproton-nucleus collisions in comparison with heavy-ion collisions due to the annihilations. Searching for the cold quark-gluon plasma (QGP) with antiproton beams has been performed as a hot topic both in experiments and in theory calculations over the past several decades. The large yields of strange particles may be produced in antiproton induced reactions, which have the advantage in comparison to proton-nucleus and heavy-ion collisions. The particle production in collisions of antiproton on nuclei has been investigated by using the intranuclear cascade (INC) model and a number of experimental data was nicely explained \cite{Cu89}. Self-consistent description of dynamical evolutions and collisions of antiproton on nucleus with transport models is still very necessary, in particular the fragmentation in collisions of antiproton on nucleus to form hypernuclei.

Strangeness production and formation of hypernuclei in antiproton induced nuclear reactions has been investigated throughly with the Giessen Boltzmann-Uehling-Uhlenbeck (GiBUU) transport model \cite{Ga12,La12}. The production of hypernuclei is associated with the reaction channels of hyperons and also hyperon-nucleon (HN) potential. From comparison of kinetic energy or momentum spectra of hyperons to experimental data, the HN potential can be extracted. Also the antinucleon-nucleon potential is able to be constrained from particle production. The dynamical mechanism on strange particle production can be explored from the analysis of reaction channels and comparison to experimental spectra.

\section{II. Model description}

In the Lanzhou quantum molecular dynamics (LQMD) model, the dynamics of resonances ($\Delta$(1232), N*(1440), N*(1535) etc), hyperons ($\Lambda$, $\Sigma$, $\Xi$) and mesons ($\pi$, $K$, $\eta$, $\overline{K}$, $\rho$, $\omega$ ) is described via hadron-hadron collisions, decays of resonances and mean-field potentials in nuclear medium \cite{Fe09,Fe11}. The evolutions of baryons (nucleons, resonances and hyperons), anti-baryons and mesons in the collisions are governed by Hamilton's equations of motion. A Skyrme-type interaction has been used in the evaluation of potential energy for nucleons and resonances.

The hyperon mean-field potential is constructed on the basis of the light-quark counting rule. The self-energies of $\Lambda$ and $\Sigma$ are assumed to be two thirds of that experienced by nucleons. And the $\Xi$ self-energy is one third of nucleon's ones. Thus, the in-medium dispersion relation reads
\begin{equation}
\omega_{B}(\textbf{p}_{i},\rho_{i})=\sqrt{(m_{B}+\Sigma_{S}^{B})^{2}+\textbf{p}_{i}^{2}} + \Sigma_{V}^{B},
\end{equation}
e.g., for hyperons $\Sigma_{S}^{\Lambda,\Sigma}= 2 \Sigma_{S}^{N}/3$, $\Sigma_{V}^{\Lambda,\Sigma}= 2 \Sigma_{V}^{N}/3$, $\Sigma_{S}^{\Xi}= \Sigma_{S}^{N}/3$ and $\Sigma_{V}^{\Xi}= \Sigma_{V}^{N}/3$.
The antibaryon energy is computed from the G-parity transformation of baryon potential as
\begin{equation}
\omega_{\overline{B}}(\textbf{p}_{i},\rho_{i})=\sqrt{(m_{\overline{B}}+\Sigma_{S}^{\overline{B}})^{2}+\textbf{p}_{i}^{2}} + \Sigma_{V}^{\overline{B}}
\end{equation}
with $\Sigma_{S}^{\overline{B}}=\Sigma_{S}^{B}$ and $\Sigma_{V}^{\overline{B}}=-\Sigma_{V}^{B}$.
The nuclear scalar $\Sigma_{S}^{N}$ and vector $\Sigma_{V}^{N}$ self-energies are computed from the well-known relativistic mean-field model with the NL3 parameter ($g_{\sigma N}^{2}$=80.82, $g_{\omega N}^{2}$=155). The optical potential of baryon or antibaryon is derived from the in-medium energy as $V_{opt}(\textbf{p},\rho)=\omega(\textbf{p},\rho)-\sqrt{\textbf{p}^{2}+m_{B}^{2}}$. A factor $\xi$ is introduced in the evaluation of antinucleon optical potential to mimic the antiproton-nucleus scattering \cite{La09} and the real part of phenomenological antinucleon-nucleon optical potential \cite{Co82} as $\Sigma_{S}^{\overline{N}}=\xi\Sigma_{S}^{N}$ and $\Sigma_{V}^{\overline{N}}=-\xi\Sigma_{V}^{N}$ with $\xi$=0.25, which leads to the optical potential $V_{\overline{N}}$=-164 MeV for an antinucleon at the zero momentum and normal nuclear density $\rho_{0}$=0.16 fm$^{-3}$. Shown in Fig. 1 is a comparison of the baryon and antibaryon energies in nuclear medium and the optical potentials as a function of baryon density. The empirical $\Lambda$ potential extracted from hypernuclei experiments \cite{Mi88} is well reproduced with the approach. The antihyperons exhibit strongly attractive potentials in nuclear medium. The optical potentials of hyperons will affect the production of hypernuclei. The values of optical potentials at saturation density are -32 MeV, -16 MeV, -164 MeV, -436 MeV and -218 MeV for $\Lambda(\Sigma)$, $\Xi$, $\overline{N}$, $\overline{\Lambda}$ and $\overline{\Xi}$, respectively.

%%%%%%%%%%%%%%%%%%%%%%%%%%%%%%%%%%%%%%% figure 1 %%%%%%%%%%%%%%%%%%%%%%%%
\begin{figure*}
\includegraphics{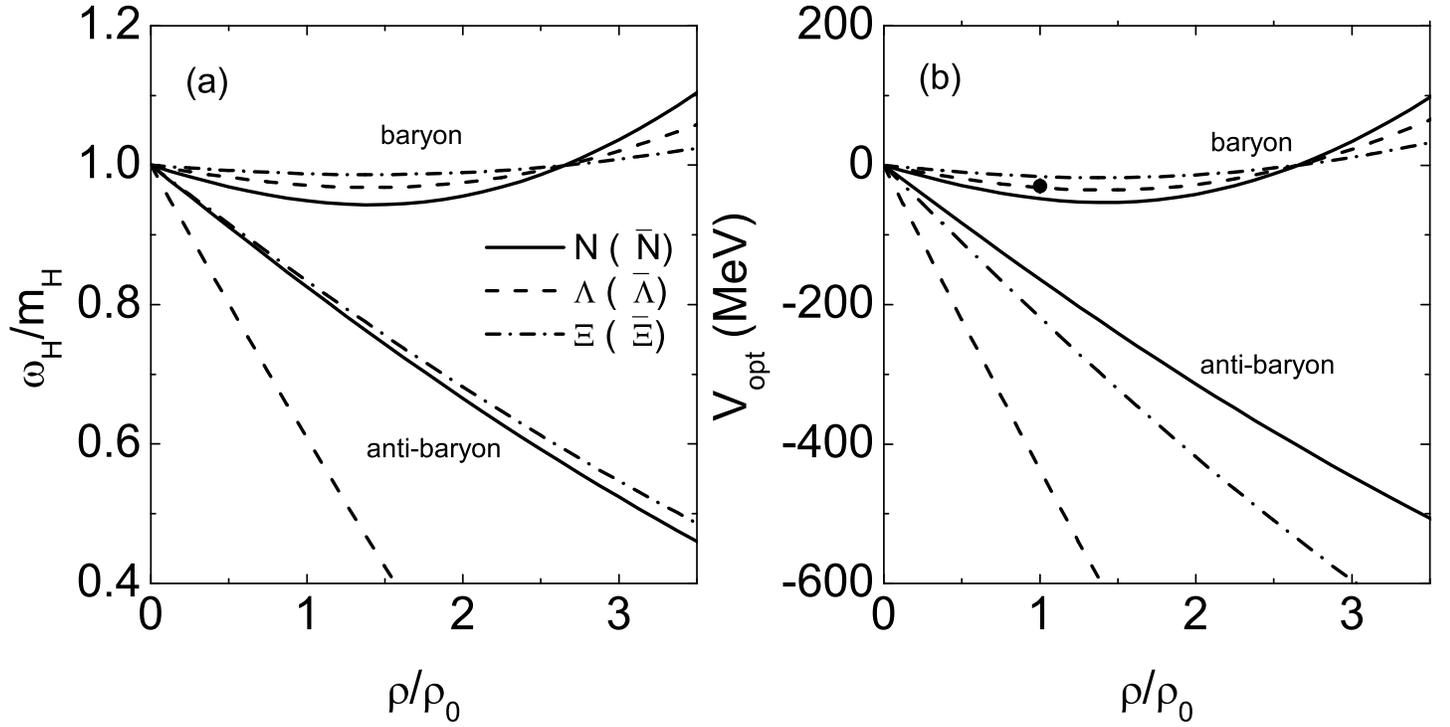}
\caption{\label{fig:wide} Density dependence of the in-medium energies of nucleon, hyperon and their antiparticles in units of free mass and the optical potentials at the momentum of $\textbf{p}$=0 GeV/c. The empirical value of $\Lambda$ potential extracted from hypernuclei experiments is denoted by the solid circle \cite{Mi88}.}
\end{figure*}
%%%%%%%%%%%%%%%%%%%%%%%%%%%%%%%%%%%%%%%%%%%%%%%%%%%%%%%%%%%%%%%%%%%%%%%%%

The Coulomb interaction between charged mesons and baryons has included in the model. The optical potentials of K and $\overline{K}$ in dense nuclear matter are considered through the dispersion relations, which are computed from the chiral Lagrangian \cite{Ka86}, but distinguishing isospin effect \cite{Fe13}. The strengths of repulsive kaon-nucleon (KN) potential and of attractive antikaon-nucleon potential with the values of 27.8 MeV and -100.3 MeV are obtained at saturation baryon density for isospin symmetric matter, respectively. The in-medium potentials of kaons and antikaons are of importance on dynamical emissions in phase space from the knowledge of heavy-ion collisions. We did not include the optical potentials for other mesons.

Based on hadron-hadron collisions in heavy-ion reactions in the region of 1-2 A GeV energies \cite{Fe11}, we have further included the annihilation channels, charge-exchange reaction, elastic and inelastic scattering in antinucleon-nucleon collisions: $\overline{N}N \rightarrow \overline{N}N$, $\overline{N}N \rightarrow \overline{B}B$,
$\overline{N}N \rightarrow \overline{Y}Y$ and $\overline{N}N \rightarrow \texttt{annihilation}(\pi,\eta,\rho,\omega,K,\overline{K},K^{\ast},\overline{K}^{\ast},\phi)$.
Here the B strands for (N, $\triangle$, N$^{\ast}$) and Y($\Lambda$, $\Sigma$, $\Xi$), K(K$^{0}$, K$^{+}$) and $\overline{K}$($\overline{K^{0}}$, K$^{-}$). The overline of B (Y) means its antiparticle. The cross sections of these channels are based on the parametrization of experimental data \cite{Bu12}. The $\overline{N}N$ annihilation is described by a statistical model with SU(3) symmetry \cite{Go92}, which includes various combinations of possible emitted mesons with the final state up to six particles \cite{La12}. A hard core scattering is assumed in two-particle collisions by Monte Carlo procedures, in which the scattering of two particles is determined by a geometrical minimum distance criterion $d\leq\sqrt{0.1\sigma_{tot}/\pi}$ fm weighted by the Pauli blocking of the final states. Here, the total cross section $\sigma_{tot}$ in mb is the sum of all possible channels. The probability reaching a channel in a collision is calculated by its contribution of the channel cross section to the total cross section as $P_{ch}=\sigma_{ch}/\sigma_{tot}$. The choice of the channel is done randomly by the weight of the probability.

\section{III. Results and discussions}

%%%%%%%%%%%%%%%%%%%%%%%%%%%%%%%%%%%%%%% figure 2 %%%%%%%%%%%%%%%%%%%%%%%%
\begin{figure*}
\includegraphics{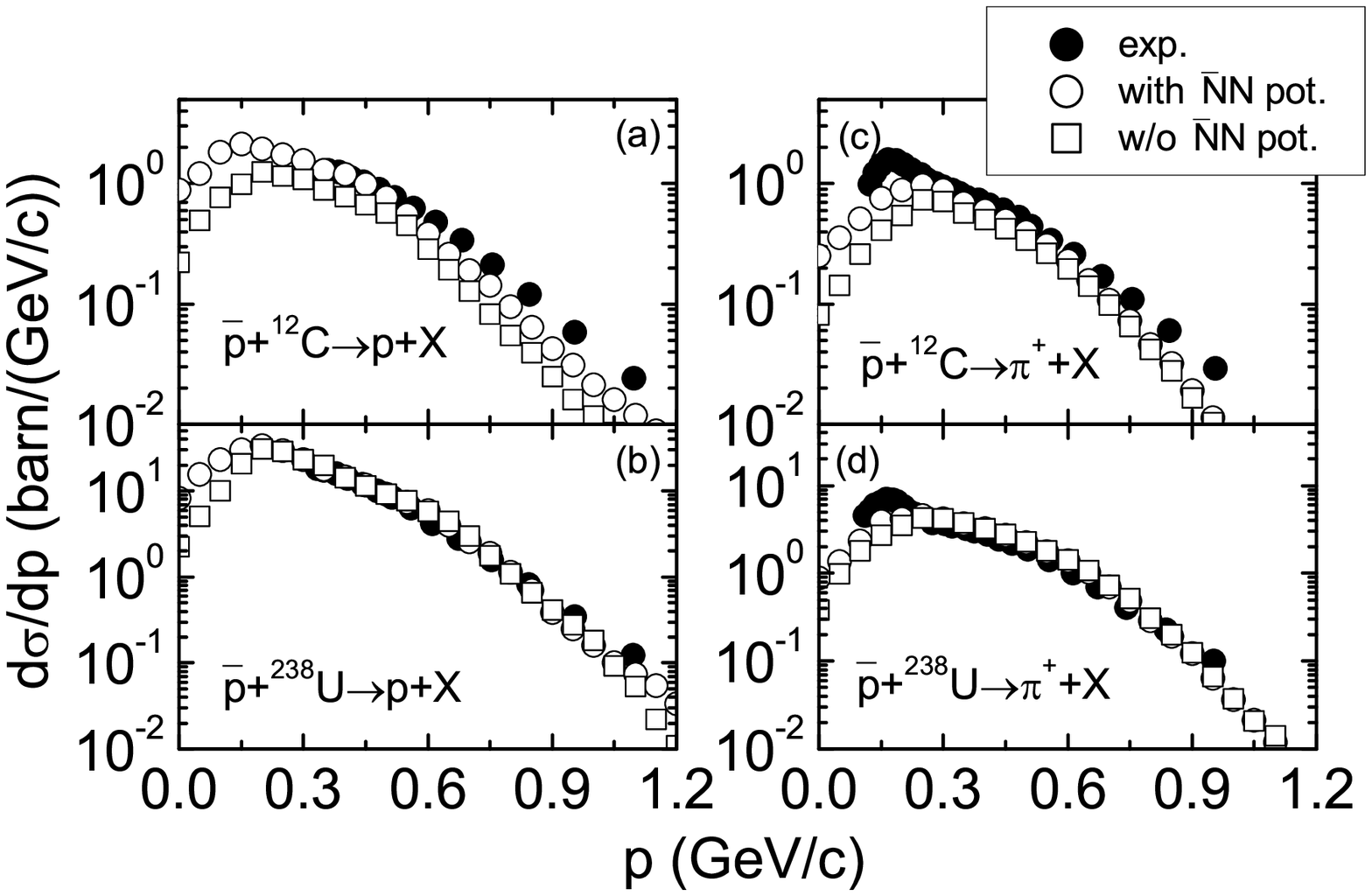}
\caption{\label{fig:wide} Momentum distributions of the angle-integrated cross sections for protons and $\pi^{+}$ produced from antiproton reactions on $^{12}$C and $^{238}$U at 608 MeV/c with and without inclusion of the $\overline{N}$N potentials. Experimental data are from Ref. \cite{Mc86}.}
\end{figure*}
%%%%%%%%%%%%%%%%%%%%%%%%%%%%%%%%%%%%%%%%%%%%%%%%%%%%%%%%%%%%%%%%%%%%%%%%%

%%%%%%%%%%%%%%%%%%%%%%%%%%%%%%%%%%%%%%% figure 3 %%%%%%%%%%%%%%%%%%%%%%%%
\begin{figure*}
\includegraphics{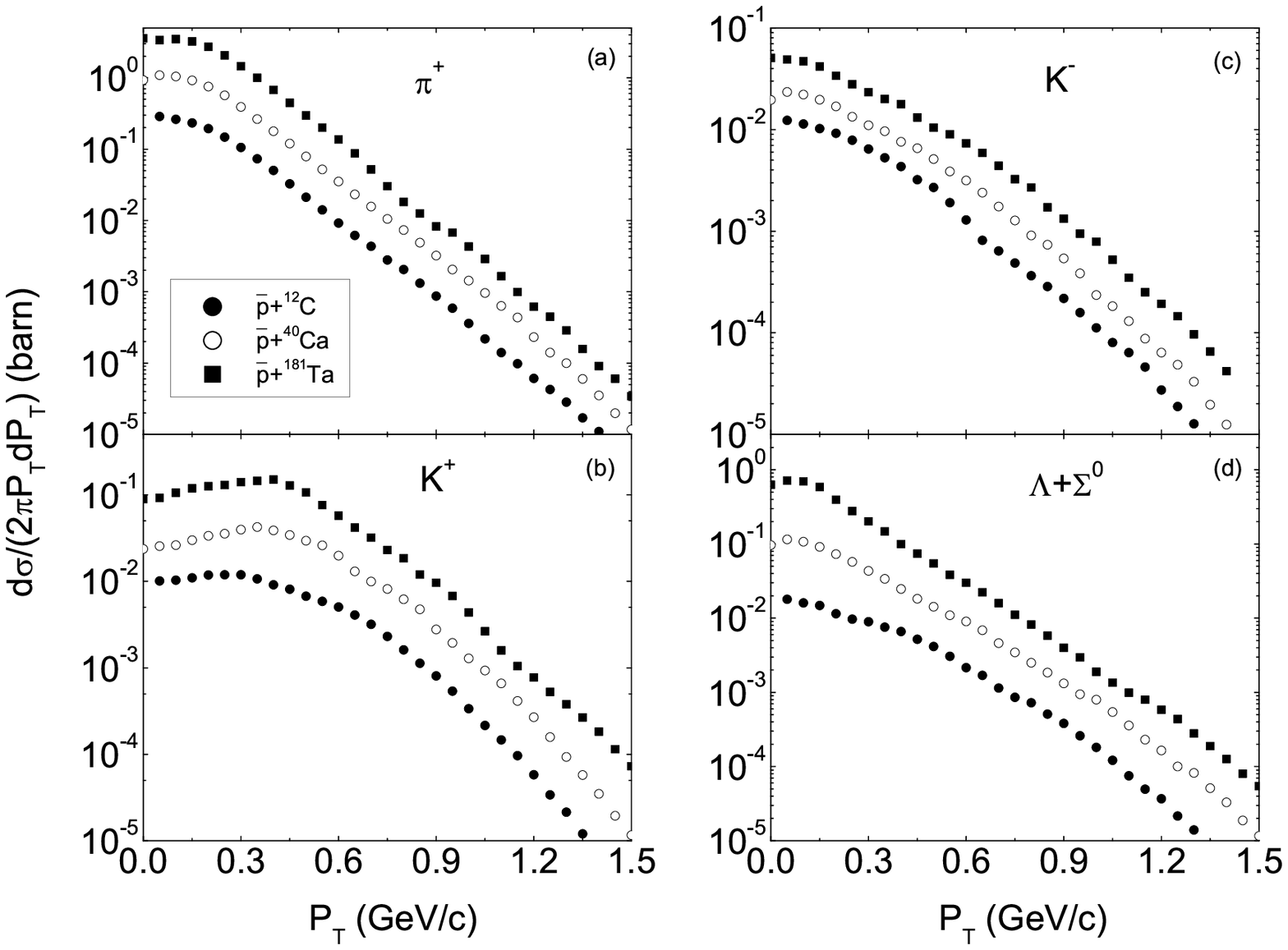}
\caption{\label{fig:wide} The transverse momentum distributions of $\pi^{+}$, K$^{+}$, K$^{-}$ and neutral hyperons produced in $\overline{p}$ on $^{12}$C, $^{40}$Ca and $^{181}$Ta at incident momentum of 4 GeV/c.}
\end{figure*}
%%%%%%%%%%%%%%%%%%%%%%%%%%%%%%%%%%%%%%%%%%%%%%%%%%%%%%%%%%%%%%%%%%%%%%%%%

Dynamics of particles produced in antiproton induced reactions can be used to extract the in-medium potentials of antiprotons and particles, in particular for strange particles in nuclear medium, which are not well understood up to now and directly affect the formation of hypernucleus. To check the reliability of the model, shown in Fig. 2 is the momentum distributions of protons and $\pi^{+}$ produced in antiproton reactions on $^{12}$C and $^{238}$U at incident momentum of 608 MeV/c. The experimental data from LEAR (Low-Energy Antiproton Ring) measurements at CERN \cite{Mc86} can be nicely reproduced with the model over the whole momentum range. The attractive $\overline{N}$N potential in nuclear medium slightly enhances the collision probabilities between antinucleons and nucleons, which leads to the increase of the momentum spectra. The shape of the proton spectra is determined by elastic collisions of antiproton on nucleons. At the considered energy, pions are mainly produced from the annihilation of $\overline{N}$N. The decay channel from $\overline{\triangle}$ formed in collisions of $\overline{p}$ on nucleons has nonnegligible contribution on the low-momentum $\pi$ production. The system size dependence of particle production in antiproton induced reactions can be observed from the transverse momentum as shown in Fig. 3. It is obvious that the yields increase with the atomic mass of target nucleus because of the reaction cross section. The distribution trends are basically the same from lighter to heavier target.

%%%%%%%%%%%%%%%%%%%%%%%%%%%%%%%%%%%%%%% figure 4 %%%%%%%%%%%%%%%%%%%%%%%%
\begin{figure*}
\includegraphics{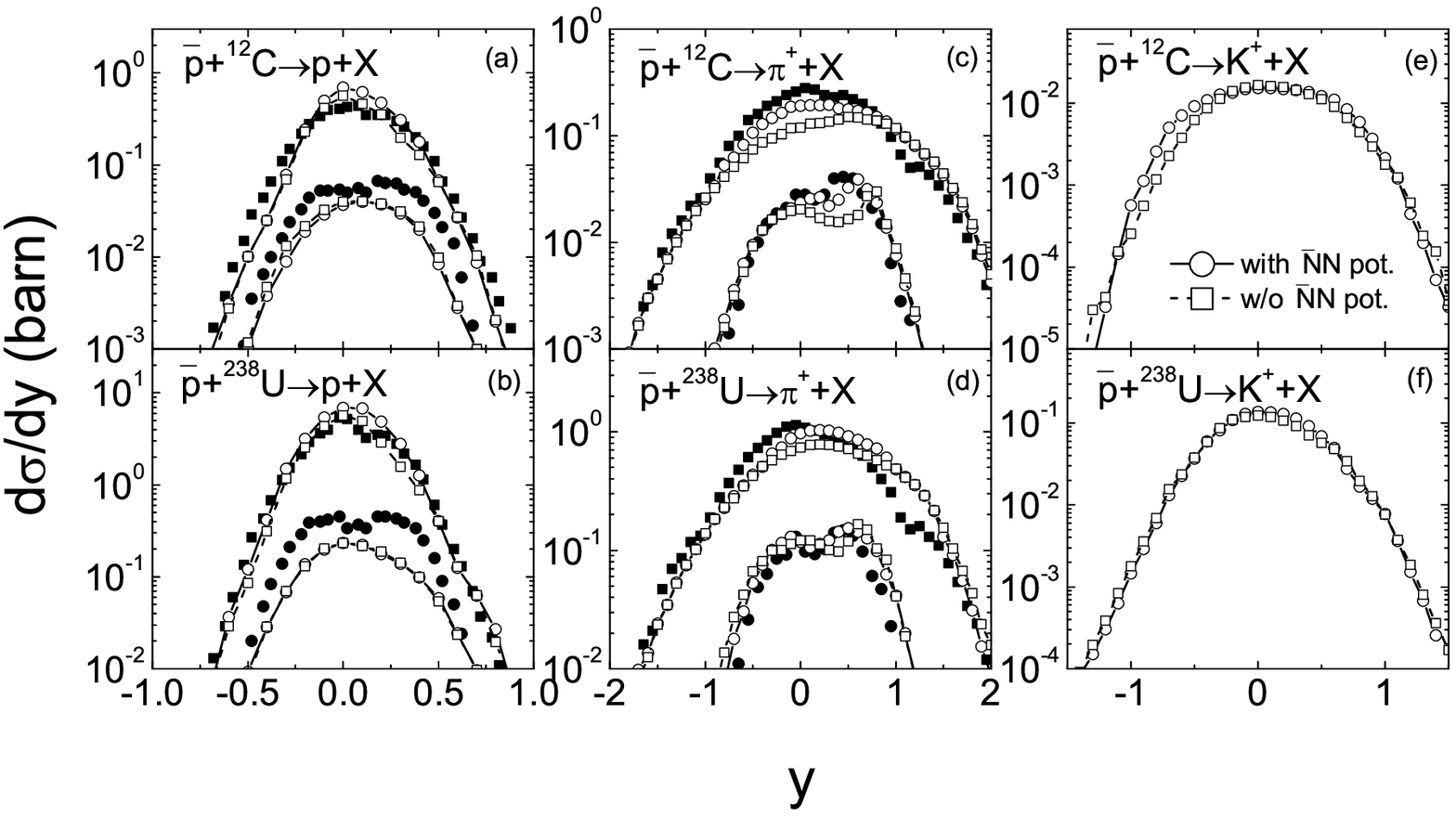}
\caption{\label{fig:wide} Rapidity distributions of protons, $\pi^{+}$ and K$^{+}$ in collisions of $\overline{p}$ on $^{12}$C and $^{238}$U at 608 MeV/c. Experimental data from LEAR measurements \cite{Mc86} are denoted by full symbols with the transverse momentum cuts $P_{T}\geq$120 MeV/c (upper spectra) and $P_{T}\geq$500 MeV/c (lower ones) for $\pi^{+}$ production, and $P_{T}\geq$330 MeV/c (upper) and $P_{T}\geq$600 MeV/c (lower) for protons.}
\end{figure*}
%%%%%%%%%%%%%%%%%%%%%%%%%%%%%%%%%%%%%%%%%%%%%%%%%%%%%%%%%%%%%%%%%%%%%%%%%

%%%%%%%%%%%%%%%%%%%%%%%%%%%%%%%%%%%%%%% figure 5 %%%%%%%%%%%%%%%%%%%%%%%%
\begin{figure*}
\includegraphics{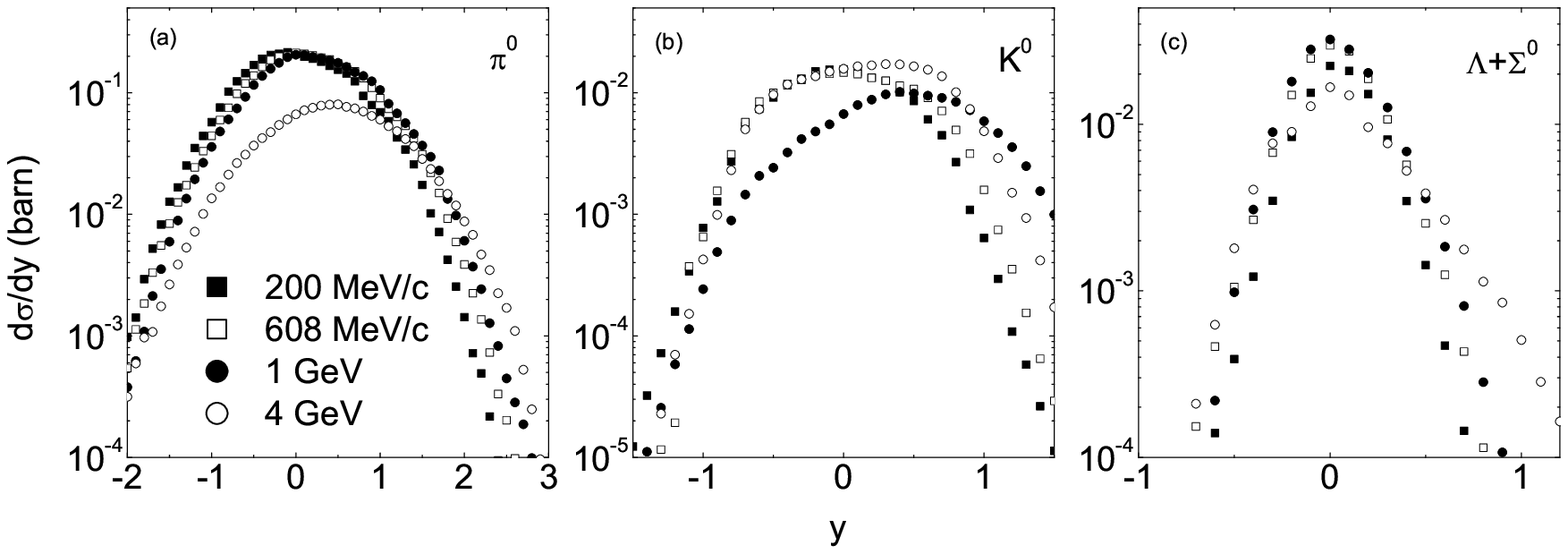}
\caption{\label{fig:wide} Rapidity distributions of $\pi^{0}$, K$^{0}$ and neutral hyperons in the reaction $\overline{p}$+$^{12}$C.}
\end{figure*}
%%%%%%%%%%%%%%%%%%%%%%%%%%%%%%%%%%%%%%%%%%%%%%%%%%%%%%%%%%%%%%%%%%%%%%%%%

The phase-space structure of particle emission in antiproton induced reactions can be also observed from the rapidity distributions. Shown in Fig. 4 is a comparison of rapidity distributions of protons, $\pi^{+}$ and K$^{+}$ in collisions of $\overline{p}$ on $^{12}$C and $^{238}$U with and without inclusion of the antinucleon-nucleon potential at incident momentum of 608 MeV/c. Experimental data from LEAR measurements \cite{Mc86} are well reproduced over the whole rapidity range. One notices that the proton emission in antiproton induced reactions almost exhibits a symmetric distribution. However, pions and kaons are produced towards the centre of mass (c.m.) rapidity of antiproton and nucleon (0.31). Influence of the $\overline{p}$N potential on particle emissions is nearly negligible besides the mid-rapidity production. Incident momentum dependence of rapidity distribution for neutral particles in the reaction of $\overline{p}$+$^{12}$C is shown in Fig. 5. The peaks of $\pi^{0}$ and K$^{0}$ move towards the c.m. rapidity of $\overline{p}$ on nucleon (y$_{\overline{p}N}$=0.11, 0.31, 0.46 and 1.08 at the incident momenta of 200 MeV/c, 608 MeV/c, 1 GeV/c and 4 GeV/c, respectively) with increasing the incident momentum. The symmetric structure appears for the neutral hyperons and the forward emission is dominant at 4 GeV/c. The results are caused from the fact that the pions and kaons are mainly produced from the annihilation of $\overline{p}$ on nucleons. The strangeness exchange reactions $\overline{K}N\rightarrow \pi Y$ contribute the hyperon production at incident momentum below the threshold value, e.g., the reaction $\overline{N}N\rightarrow \overline{\Lambda}\Lambda$ (p$_{threshold}$=1.439 GeV/c). The hyperon is also produced via meson induced reactions $M N\rightarrow KY$ with M being $\pi,\eta,\rho,\omega$.

%%%%%%%%%%%%%%%%%%%%%%%%%%%%%%%%%%%%%%% figure 6 %%%%%%%%%%%%%%%%%%%%%%%%
\begin{figure*}
\includegraphics{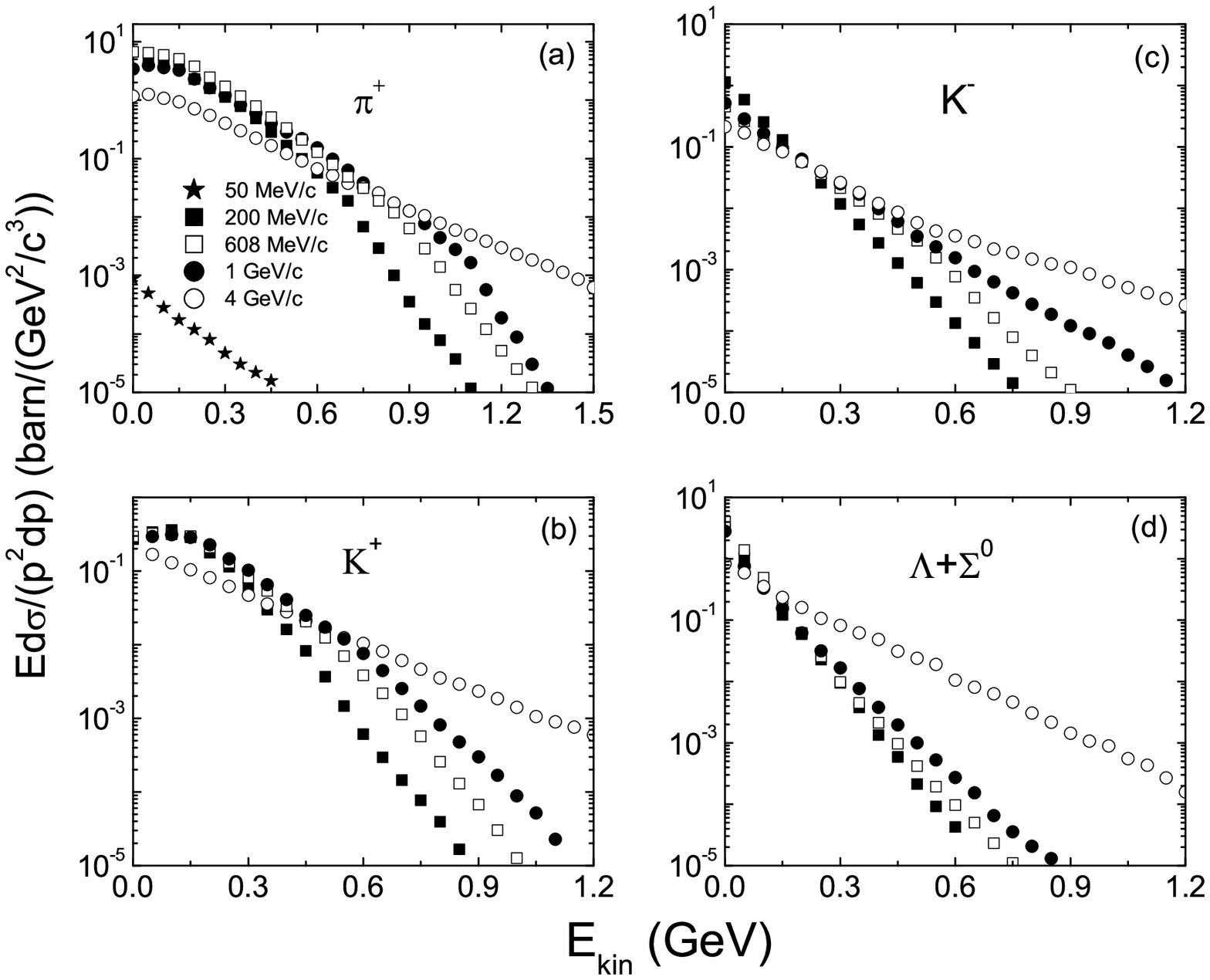}
\caption{\label{fig:wide} The kinetic energy spectra of inclusive invariant cross sections for the production of $\pi^{+}$, K$^{+}$, K$^{-}$ and neutral hyperons in collisions of $\overline{p}$ on $^{12}$C at incident momenta of 50 MeV/c, 200 MeV/c, 608 MeV/c, 1 GeV/c and 4 GeV/c, respectively.}
\end{figure*}
%%%%%%%%%%%%%%%%%%%%%%%%%%%%%%%%%%%%%%%%%%%%%%%%%%%%%%%%%%%%%%%%%%%%%%%%%

%%%%%%%%%%%%%%%%%%%%%%%%%%%%%%%%%%%%%%% figure 7 %%%%%%%%%%%%%%%%%%%%%%%%
\begin{figure*}
\includegraphics{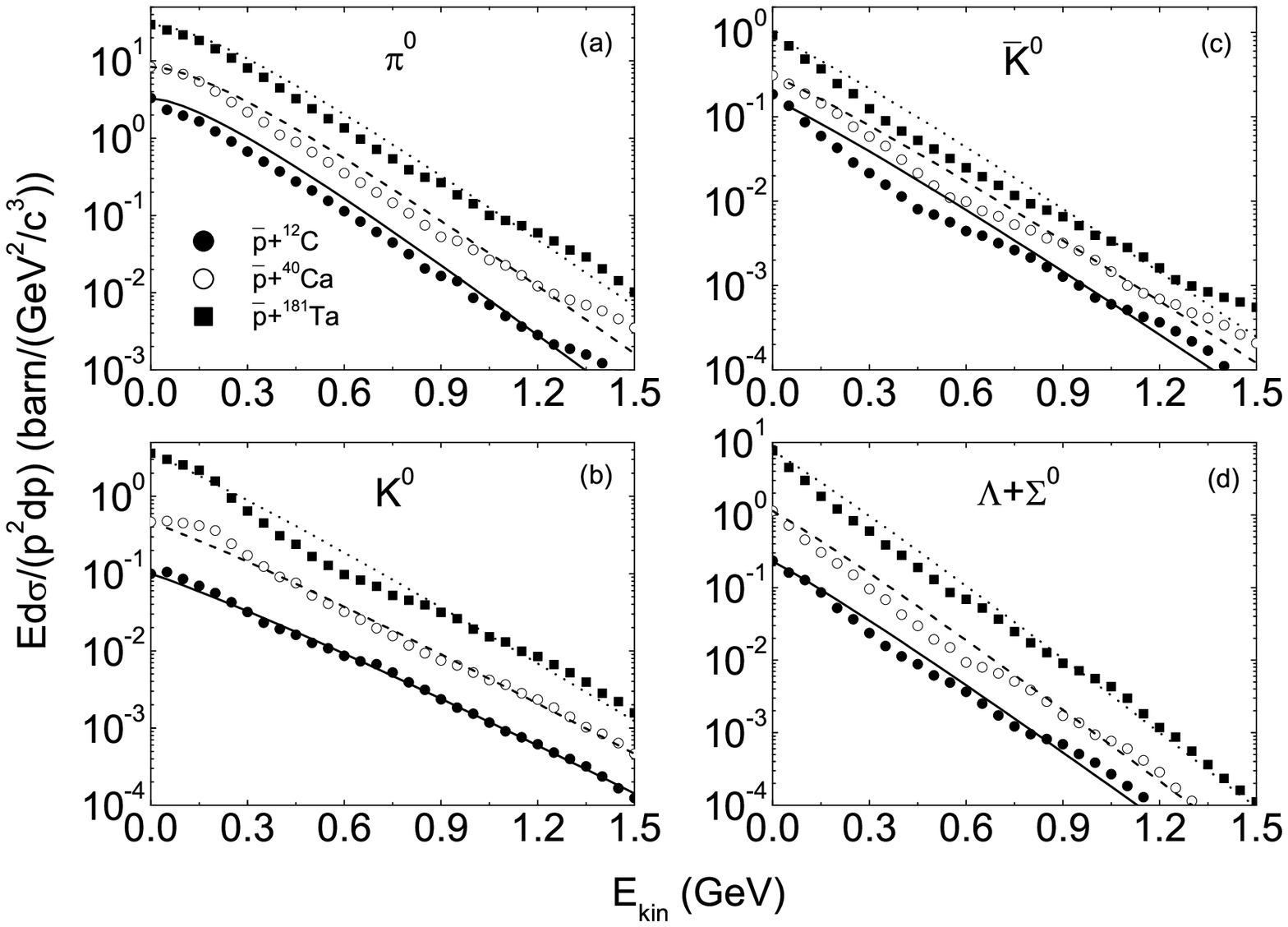}
\caption{\label{fig:wide} The inclusive spectra of neutral particles produced in the reaction of $\overline{p}$ on $^{12}$C, $^{40}$Ca and $^{181}$Ta at 4 GeV/c.}
\end{figure*}
%%%%%%%%%%%%%%%%%%%%%%%%%%%%%%%%%%%%%%%%%%%%%%%%%%%%%%%%%%%%%%%%%%%%%%%%%

More information of particle production in antiproton induced reactions can be obtained from the invariant spectra. Shown in Fig. 6 is the kinetic energy spectra of inclusive invariant cross sections for the production of $\pi^{+}$, K$^{+}$, K$^{-}$ and neutral hyperons in collisions of $\overline{p}$ on $^{12}$C at incident momenta of 50 MeV/c, 200 MeV/c, 608 MeV/c, 1 GeV/c and 4 GeV/c, respectively. It can be seen that the spectra become more and more flat with increasing the incident momentum, in particular for neutral hyperons at above the threshold momentum. Dependence of the spectra on reaction system is also investigated for neutral particles at the incident momentum of 4 GeV/c in Fig. 7. It is obvious the production yields is increasing with heavier target because of the larger geometric cross sections. The invariant cross section can be fitted by the Maxwell-Boltzmann distribution as
\begin{equation}
\frac{Ed\sigma}{p^{2}dp}=C E\exp(-E_{kin}/T),
\end{equation}
where $E$ and $E_{kin}$ are the total energy and the kinetic energy, respectively, and the normalization $C$ and temperature $T$ being the fitting parameters. The inverse slope parameters $T$ have been obtained for $\pi^{0}$ being the values of 103$\pm$4 MeV, 106$\pm$3 MeV and 110$\pm$2 MeV for $^{12}$C, $^{40}$Ca and $^{181}$Ta, respectively. The temperature parameters of 150$\pm$7 MeV, 144$\pm$4 MeV and 136$\pm$3 MeV for K$^{0}$, 124$\pm$16 MeV, 130$\pm$10 MeV and 124$\pm$7 MeV for $\overline{K}^{0}$, 101$\pm$16 MeV, 97$\pm$15 MeV and 94$\pm$8 MeV for neutral hyperons are obtained for $^{12}$C, $^{40}$Ca and $^{181}$Ta, respectively. The results are basically consistent with the experimental data by KEK in $\overline{p}$+$^{181}$Ta reaction at incident momentum of 4 GeV/c \cite{My88}, where the temperatures of 135$\pm$13 MeV and 97$\pm$6 MeV were obtained for K$^{0}_{S}$ and $\Lambda$, respectively. The conclusions are also useful for the future PANDA experiments. The lower temperature parameters are obtained for pions, antikaons and hyperons in comparison to the ones of kaons, which are caused from the secondary collisions, e.g., $\pi N\leftrightarrow \Delta$, $\pi N\rightarrow KY$, $\overline{K}N\leftrightarrow \pi Y$.

%%%%%%%%%%%%%%%%%%%%%%%%%%%%%%%%%%%%%%% figure 8 %%%%%%%%%%%%%%%%%%%%%%%%
\begin{figure*}
\includegraphics{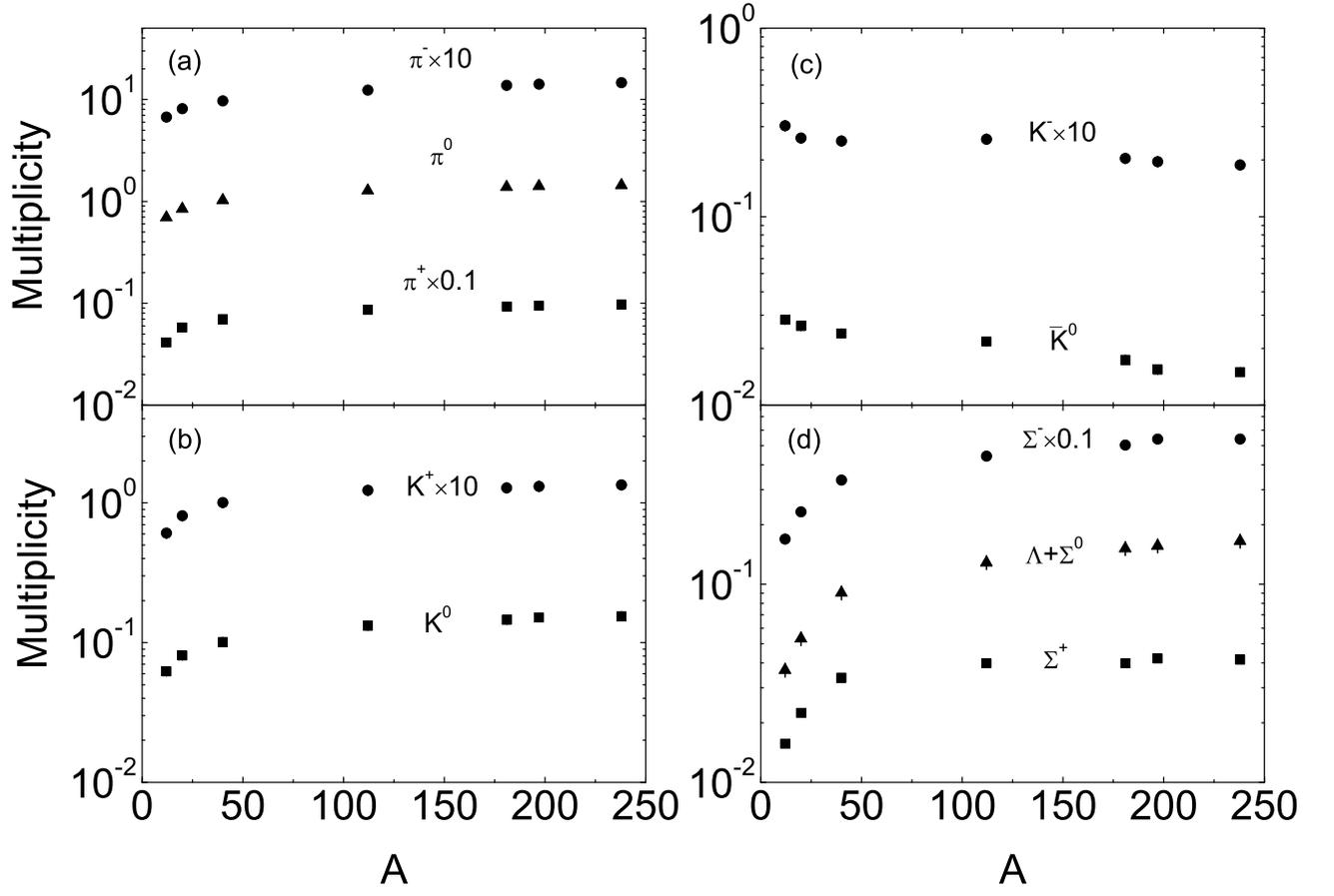}
\caption{\label{fig:wide} Total multiplicities of particles as a function of atomic number of target nuclei in antiproton induced reactions on $^{12}$C, $^{20}$Ne, $^{40}$Ca, $^{112}$Sn, $^{181}$Ta, $^{197}$Au and $^{238}$U at 4 GeV/c.}
\end{figure*}
%%%%%%%%%%%%%%%%%%%%%%%%%%%%%%%%%%%%%%%%%%%%%%%%%%%%%%%%%%%%%%%%%%%%%%%%%

To avoid the geometric effects on particle production in antiproton induced reactions, the multiplicities of pions, kaons, antikaons and hyperons are calculated from the carbon to uranium in Fig. 8. It is obvious that a flat structure appears in the production of pions and kaons. The number of antikaons decreases with the atomic mass of target nuclide. However, the hyperon production gives an opposite trend because of the strangeness exchange reactions. We have included the hyperon-nucleon potential in the calculation. It would be a nice approach with medium or heavy target to produce hypernucleus in the antiproton induced reactions. The work on hypernucleus production in the antiproton induced reactions with the LQMD transport model combined with the statistical model is in progress.

\section{IV. Conclusions}

The LQMD model has been further improved to investigate the nuclear dynamics induced by antiprotons. Dynamics on particle production, in particular pions, kaons, antikaons and hyperons, has been investigated within the model. The $\overline{N}$N potential slightly changes the structures of rapidity and momentum distributions of $\pi^{+}$ and protons in comparison to the LEAR data. The yields of pions, kaons and antikaons are mainly contributed from the annihilations of antiproton on nucleons. Hyperons are dominated via the meson-nucleon collisions and strangeness exchange reactions when the incident momentum is below the annihilation threshold value. The obtained nuclear temperature is still below the condition of QGP formation.

The hyperon-nucleon potential has been shown to be importance of hyperon emissions in phase space in heavy-ion collisions \cite{Fe13}. The evaluation of hyperons captured by fragments to form hypernuclei
will be helpful for measurements in experiments. Calculations on dynamical distributions of hypernuclei in phase space is in progress, in particular for the hypernuclei of s=-2 and s=1 ($_{\Lambda\Lambda}X$, $_{\Xi}X$ and $_{\overline{\Lambda}}X$ with $X$ being fragment).

\textbf{Acknowledgements}

This work was supported by the National Natural Science Foundation of China Projects (Nos 11175218 and U1332207) and the Youth Innovation Promotion Association of Chinese Academy of Sciences. The author (ZQF) is grateful to the support of K. C. Wong Education Foundation (KCWEF) and DAAD during his research stay in Justus-Liebig-Universit\"{a}t Giessen, Germany.

\end{document}